\documentclass[aps,prb,12pt,amsmath,amssymb,tightenlines]{revtex4}

\def\xx{\bm x} \def\yy{\bm y} \def\pp{\bm p} \def\uu{\bm u} \def\aa{\bm a} \def\AA{{\bm A}}
\def\dxx{ d^3\bm x } \def\epsovermh{{\frac{\epsilon}{m\hbar}}}
\font\bbtwelve=msbm10 scaled\magstep1 

\usepackage{bm}

\begin{document}

\hfill\hbox{\small To appear in the \textit{Amer. J. Phys.}, 2004} \vskip 2pt

\title{Path integral in a magnetic field using the Trotter product formula}

\author{B. Gaveau}
\email{gaveau@ccr.jussieu.fr}
\affiliation{Laboratoire Analyse et Physique Math\'ematique, 14 Avenue F\'elix Faure, 75015 Paris, France}

\author{E. Mih\'okov\'a}
\email{mihokova@fzu.cz}
\affiliation{Institute of Physics, Academy of Sciences of the Czech Republic, Cukrovarnick\'a~10, 162~53~Prague~6,
Czech~Republic}

\author{M. Roncadelli}
\email{marco.roncadelli@pv.infn.it}
\affiliation{Istituto Nazionale di Fisica Nucleare, Sezione di Pavia, 27100 Pavia, Italy}

\author{L. S. Schulman}
\email{schulman@clarkson.edu}
\affiliation{Physics Department, Clarkson University, Potsdam, New York 13699-5820}

\begin{abstract}
The derivation of the Feynman path integral based on the Trotter product formula is extended to the case where the system is in a magnetic field.
\end{abstract}

\maketitle

\section{\label{sec:intro}Introduction}
There are many ways to derive the path integral. In his original paper, Feynman\cite{feynman} gave a physical motivation for a particular form and verified its correctness by propagating the wave function for an infinitesimal time step. In Ref.~\onlinecite{lsspibook} the opposite perspective is used: start with a Hamiltonian and use the Trotter product formula\cite{nelson} to arrive at the sum-over-paths. Some of Feynman's poetry is lost, but one connects more easily to familiar formulations. The Trotter-product-formula-based derivation in Ref.~\onlinecite{lsspibook} is limited to Hamiltonians of the form $H={\bm p}^2/2m+V({\bm x})$, and skirts the more delicate case of a particle in a magnetic field. As is well known, the latter situation requires that when writing the classical action for the broken line path in the path integral, the vector potential, $\bm A(\bm x)$, be approximated by its average at the ends of each straight segment of the path. One can also evaluate $\bm A$ at the midpoint of that segment, giving rise to the ``midpoint rule.'' This feature is one of the fundamental properties of the path integral, reflecting the close relation between the path integral and the Wiener integral, and the ``infinite velocities'' or non-differentiability of the paths in both cases. Yet more delicate evaluations are necessary for the path integral on curved spaces, so it would be useful to understand how to deal with these rough paths from the Trotter derivation perspective.

In this article we use a slight variation of the derivation in Ref.~\onlinecite{lsspibook} to show how we can arrive at the correct path integral, including a magnetic field. Some years ago two of us published such a derivation,\cite{sensitive} but it was phrased in the language of stochastic processes. The present work uses a framework more familiar to physicists and is more streamlined. The earlier paper includes path integrals on curved spaces, which are not treated here. Because the purpose of this article is to extend familiar path integral techniques, we do not elaborate on how the formulae developed here make their way into the usual path integral representations.

\section{\label{sec:propagator}Breaking up the propagator}
The propagator is the operator $\exp(-iHt/\hbar)$, with $H$ the Hamiltonian. $H$ is taken to be
\begin{equation}
H=\frac1{2m} (\bm p -\frac ec\bm A(\bm x))^2 +V(\bm
x),
\label{eq:H}
\end{equation}
in the usual notation. For $\bm a,\bm b \in \hbox{\bbtwelve R}^3$, the propagator is exactly given by
\begin{eqnarray}
G(\bm b,t;\bm a)&=&\langle \bm b|\exp(-iHt/\hbar)|\bm a\rangle \\
 &=&\langle \bm b|\big[\exp(-iHt/N\hbar)\big]^N|\bm a\rangle \\
 &=& \langle \bm b|\exp(-iHt/N\hbar)
 \int\dxx_1 |\xx_1\rangle\langle\xx_1|
 \exp(-iHt/N\hbar) \nonumber\\
 {}&&\times\int\dxx_2 |\xx_2\rangle\langle\xx_2|
 \exp(-iHt/N\hbar) \times \ldots \nonumber\\
 {}&& \times \int\dxx_{N-1}
|\xx_{N-1}\rangle\langle\xx_{N-1}|
 \exp(-iHt/N\hbar)
 |\bm a\rangle.
\label{eq:powers}
\end{eqnarray}
Equation~(\ref{eq:powers}) can be written more concisely as
\begin{equation}
G(\bm b,t;\bm a)=\!\int \prod_{k=1}^{N-1}\dxx_k 
\prod_{\ell=0}^{N-1} G(\xx_{\ell+1},\epsilon;\xx_{\ell}) ,
\label{eq:niceproduct}
\end{equation}
with $\epsilon=t/N$, $\xx_N=\bm b$, and $\xx_0=\bm a$.

The basic fact of life about the short time propagators, $G(\xx_{\ell+1},\epsilon;\xx_{\ell})$, is that for $N\to\infty$, we need only maintain O($\epsilon$) accuracy. That is, they can be replaced by other objects that differ only by terms that go to zero faster than $\epsilon$. For ordinary numbers this can be seen by recalling that $e^x=\lim_{N\to\infty}(1+\frac xN)^N$, and by observing that adding to $x$ any quantity that goes to zero with $N$ does not change the result. That is, $(1+\frac{x+a_N}{N})^N\to e^x$ provided $a_N\to0$. To show that this limit works for unbounded operators is of course a much bigger project and the reader is referred to Refs.~\onlinecite{lsspibook} and \onlinecite{nelson} and to references therein.

Our goal therefore is to approximate $G(\xx,\epsilon;\yy)$ to first order in $\epsilon$. We return to operator language. The starting point is to observe that for two operators
\begin{equation}
\exp[\lambda(A+B)]=  \exp(\lambda A)\exp(\lambda B)\exp \big(\frac{\lambda^2}2[B,A]+\hbox{O}(\lambda^3)\big).
\label{eq:trotterzero}
\end{equation}
The propagator is of this form, with $\lambda=\epsilon$, $A=-iK/\hbar$, $B=-iV/\hbar$, and $K$ the kinetic energy. We can also write Eq.~(\ref{eq:trotterzero}) as\cite{exponentisbetter}
\begin{equation}
\exp[\lambda(A+B)]=  \exp(\lambda A)\exp(\lambda B) + \hbox{O}(\lambda^2).
\label{eq:trotter}
\end{equation}
This formula lets us replace $\exp[\lambda(A+B)]$ by $\exp(\lambda A)\exp(\lambda B)$ when $\lambda$ is $\epsilon$.

In the absence of a vector potential one term is a function of momentum only and the other a function of position only, allowing the breakup that leads to the usual path integral. For later reference we review this procedure. The potential, $V$, is diagonal in position space, so that we immediately obtain
\begin{equation}
G(\xx,\epsilon;\yy)= \langle\xx| \exp(-iK\epsilon/\hbar) |\yy\rangle
 \exp(-iV(\yy)\epsilon/\hbar) + \hbox{O}(\epsilon^2).
\label{eq:Gone}
\end{equation}
If $K$ is simply $\bm p^2/2m$, the next step is straightforward. By inserting $\int d^3\pp\,|\pp\rangle\langle \pp|$ to the left of $|\yy\rangle$, we find that $\exp(-iK\epsilon/\hbar)$ gives\cite{operatorhat}
\begin{equation}
\int\!d^3\bm p\langle\xx|\exp(-i\hat{\bm p}^2\epsilon/\hbar2m)
|\bm p\rangle\langle \bm p|\yy\rangle = \big(\frac m {2\pi
i\hbar\epsilon}\big)^{3/2}
\exp\Big(\frac i\hbar \frac m2\frac{(\xx-\yy)^2}\epsilon\Big).
\label{eq:Gtwo}
\end{equation}
(See Eq.~(\ref{eq:momentumeigenstate}) for the form of the momentum eigenstates.) If we combine Eqs.~(\ref{eq:Gone}) and (\ref{eq:Gtwo}), we obtain the expression, which when iterated, gives the path integral in three dimensions:
\begin{equation}
G(\xx,\epsilon;\yy)= \big(\frac m {2\pi i\hbar\epsilon}\big)^{3/2}
   \exp\Big[\frac i\hbar \big(\frac{m(\xx-\yy)^2}{2\epsilon}-\epsilon
   V(\yy)\big)\Big] + \hbox{O}(\epsilon^2).
\label{eq:Gthree}
\end{equation}
Thus, if we insert Eq.~(\ref{eq:Gthree}) in Eq.~(\ref{eq:niceproduct}), we obtain the classical action in the exponent. Note that by interchanging $K$ and $V$ in Eq.~(\ref{eq:trotter}), the argument of $V$ in Eq.~(\ref{eq:Gthree}) becomes $\xx$ rather than $\yy$. As indicated, this changes the short-time propagator by less than O($\epsilon$) and therefore does not change the final result. Indeed, any appropriately weighted sum of $V(\xx)$ and $V(\yy)$ can be used, or as we shall see below, $V$ can be taken at any point on the line between $\xx$ and $\yy$.

\section{\label{sec:uncomplete}Uncompleting the square}
When a magnetic field is present, the expression $\exp(-iK\epsilon/\hbar)$ is more complicated, because $K\equiv [ \bm p -e\bm A(\bm x)/c]^2/2m$. Under these circumstances, inserting a momentum-state resolution of the identity is inadequate, because $\bm A$ is a function of $\xx$. The idea of the next step is to look at the square root of $K$, which because it is a sum of $\bm p$ and $\bm A$, can be resolved by separately expressing them in momentum and position space bases. Taking the square root in the exponent is accomplished at the expense of introducing an additional integral, a process variously known as ``uncompleting the square''\cite{kac} or the ``Gaussian trick'' \cite{amit}. The idea is based on the identity
\begin{equation}
 \exp\big(\frac{\bm b^2}{2}\big)=
 \big( \frac1{\sqrt{2\pi}} \big)^{3}\!\int\! d^3\bm u\, 
 \exp\big(-\frac{\bm u^2}{2} +\bm b\cdot \bm u \big).
\label{eq:uncompletezero}
\end{equation}
We use the following variant of Eq.~(\ref{eq:uncompletezero}):
\begin{equation}
 \exp\big(-i\epsilon\frac{\bm b^2}{2m\hbar}\big)=
 \big( \frac1{\sqrt{2\pi i}} \big)^{3}
 \!\int\! d^3\bm u\, 
 \exp\big(i\frac{\bm u^2}{2}
-i\sqrt{\frac{\epsilon}{m\hbar}}\,\bm b\cdot \bm u 
\big).
\label{eq:uncomplete}
\end{equation}
For convenience, we let $\bm a(\bm x)\equiv e\bm A(\bm x)/c$. We want to evaluate Eq.~(\ref{eq:Gone}) and concentrate on the kinetic energy part, which, with the aid of Eq.~(\ref{eq:uncomplete}) is rewritten as
\begin{eqnarray} 
 G_K(\xx,\epsilon;\yy)&\equiv& \langle\xx| \exp(-iK\epsilon/\hbar) |\yy\rangle \nonumber\\
  &=& \langle\xx| \exp\big(-i\epsilon \frac{(\bm p - \bm a)^2}{2m\hbar}\big) |\yy\rangle   \nonumber\\
  &=& \big( \frac1{\sqrt{2\pi i}} \big)^{3} \!\int\! d^3\bm u\, 
     \exp\big(i\frac{\bm u^2}{2}\big)
    \langle\xx| \exp\big(-i\sqrt{\frac{\epsilon}{m\hbar}}\,(\bm p - \bm a)\cdot \bm u \big)  |\yy\rangle.
\label{eq:linear}
\end{eqnarray}
Equation~(\ref{eq:linear}) shows where the trouble enters. The $\bm p$ has been separated from the $\bm a$ with which it does not commute. That's the good part. Nevertheless, Eq.~(\ref{eq:trotter}) would not provide the needed accuracy, because it is the \textit{square root} of $\epsilon$ that appears in Eq.~(\ref{eq:linear}) ($\lambda$ of Eq.~(\ref{eq:trotter}) is now $\sqrt{\epsilon}$). The error is the square of this quantity, namely $\epsilon$ itself, which cannot be neglected. This problem will occur whenever an expression mixes non-commuting variables, for example in a curved-space metric (or kinetic energy). Generally speaking, this is why the operator ordering problem does not get solved in the path integral. The way to deal with this problem is to improve on Eq.~(\ref{eq:trotter}).  To this we now turn.

\section{\label{sec:accurate}A more accurate product formula}
A slight variation of Eq.~(\ref{eq:trotter}) provides O($\lambda^3$) accuracy:
\begin{equation}
\exp[\lambda(A+B)]=  \exp(\lambda B/2)\exp(\lambda A)\exp(\lambda B/2)
        +\hbox{O}(\lambda^3).
\label{eq:betterproduct}
\end{equation}
Equation~(\ref{eq:betterproduct}) can be checked by direct expansion of the exponents.\cite{anotherproof} A different symmetrization was used in Ref.~\onlinecite{sensitive}. Equation~(\ref{eq:betterproduct}) has also been used for increased accuracy in the numerical evaluation of path integrals.\cite{janke}

We apply Eq.~(\ref{eq:betterproduct}) to $\langle\xx| \exp\big( -i\sqrt{\epsilon}(\bm p - \bm a)\cdot \bm u/\sqrt{m\hbar} \big) |\yy\rangle$ to obtain
\begin{eqnarray}
&&\langle\xx| \exp\big( -i\sqrt{\epsovermh} (\bm p - \bm a)\cdot
\bm u \big)
 |\yy\rangle \nonumber\\
 &&\quad=
 \langle\xx| \exp\big( i\sqrt{\epsovermh} \bm a \cdot
\bm u/2 \big)
\nonumber\\
 {}&& \qquad \times \exp\big( -i\sqrt{\epsovermh}\bm p\cdot
\bm u \big)
 \exp\big( i\sqrt{\epsovermh} \bm a\cdot \bm u/2 \big)
 |\yy\rangle \nonumber\\
 &&\quad=
 \exp\big( i\sqrt{\epsovermh} \bm a(\xx) \cdot \bm
u/2 \big) \nonumber\\
{}&& \qquad\times \langle\xx| 
 \exp\big( -i\sqrt{\epsovermh}\bm p\cdot \bm u \big)
 |\yy\rangle
 \exp\big( i\sqrt{\epsovermh} \bm a(\yy)\cdot \bm u/2 
\big).
\label{eq:Gfour}
\end{eqnarray}
The error in the first ``equality'' is O($\epsilon^{3/2}$), although it is not explicitly indicated. Once again, a momentum resolution of the identity is inserted in the $\pp$-dependent portion of Eq.~(\ref{eq:Gfour}):
\begin{equation}
\langle\xx| \exp\big( -i\sqrt{\epsovermh}\hat{\bm p}\cdot \bm u \big)
 |\yy\rangle =\!\int\! d^3\pp\,\langle\xx| \exp\big( -i\sqrt{\epsovermh}\hat{\bm p}\cdot \bm u 
 \big)| \pp\rangle\langle\pp|\yy\rangle.
\label{eq:Gfive}
\end{equation}
If we recall that
\begin{equation}
\langle \xx|\pp\rangle = ({2\pi\hbar})^{-3/2}\exp(i\pp\cdot\xx/\hbar),
\label{eq:momentumeigenstate}
\end{equation}
then Eq.~(\ref{eq:Gfive}) becomes
\begin{eqnarray}
\langle\xx| \exp( -i\sqrt{\epsilon}\bm p\cdot \bm u/\sqrt{m\hbar})
 |\yy\rangle &=&\Big(\frac1{2\pi\hbar}\Big)^{3} \!\int\! d^3\pp\,
 \exp ( -i\sqrt{\epsilon}{\bm p}\cdot \bm u/\sqrt{m\hbar}+i\pp\cdot(\xx-\yy)/\hbar)
\nonumber\\ 
&=&\frac1{\hbar^3} \delta^3 ( -\sqrt{\epsilon}\bm u/\sqrt{m\hbar}+(\xx-\yy)/\hbar) \nonumber\\ 
&=&\frac1{\hbar^3} \Big[\sqrt{\frac{m\hbar}\epsilon}\Big]^3  \delta^3 (-\bm u+(\xx-\yy)\sqrt{m/\hbar\epsilon} ).
\label{eq:Gsix}
\end{eqnarray}
The integral over $\bm u$ in Eq.~(\ref{eq:linear}) is now trivial, yielding for $G_K$,
\begin{eqnarray}
G_K(\xx,\epsilon;\yy)&=&
 \big[\frac{1}{2\pi
i} \frac{m}{\epsilon\hbar}\big]^{\frac32}
 \!\int\! d^3\bm u\, e^{i\uu^2/2}
\exp\bigg[i
\sqrt{\epsovermh}\uu\cdot\big(\frac{\aa(\xx)+\aa(\yy)}{2}\big) 
 \bigg] \delta^3\Big(
 \frac {(\xx-\yy)} {\sqrt{\hbar\epsilon/m}}-\bm u
 \Big) \nonumber\\
 &=&\big[\frac{m}{2\pi i\hbar\epsilon}\big]^{\frac32}
 \exp\big[ \frac i \hbar \frac
m2\frac{\big(\xx-\yy\big)^2}{\epsilon} \big]
 \exp\big[\frac i \hbar \big(\xx-\yy\big) \cdot
\big(\frac{\aa(\xx)+\aa(\yy)}{2}\big) 
 \big],
\label{eq:GsubK}
\end{eqnarray}
with $\aa\equiv e\bm A/c$. For the classical Lagrangian, a magnetic field contributes $\bm v\cdot e\bm A(\xx)/c$. If we multiply and divide by $\epsilon$ in the last expression in Eq.~(\ref{eq:GsubK}), we obtain exactly the appropriate term. For completeness, we restore $\bm A$ and $V$, yielding,
\begin{eqnarray}
G(\xx,\epsilon;\yy)&=&
\big[\frac{m}{2\pi i\hbar\epsilon}\big]^{\frac32}
 \exp\bigg\{\frac i \hbar \epsilon \big[ \frac
m2\frac{\big(\xx-\yy\big)^2}{\epsilon^2}
 +\frac{\big(\xx-\yy\big)}\epsilon \cdot
 \frac ec \big(\frac{\AA(\xx)+\AA(\yy)}{2}\big) -V(\yy)
 \big]\bigg\} \nonumber\\
 &&~~~~~~~~~~~~~~~ +\hbox{O}(\epsilon^{3/2}) \,.
\label{eq:Gfull}
\end{eqnarray}
The argument of the exponent is seen to be $i\epsilon/\hbar$ times the classical Lagrangian.

Note that what naturally arises is the average of the vector potential, $\bm A$, at the endpoints of the broken line path. The ``midpoint'' way of attaining the same level of precision uses $\bm A[(\xx+\yy)/2]$. The difference between these forms is essentially a second derivative of $\bm A$ times $(\xx-\yy)^2$. The latter is of order $\epsilon$ and in turn multiplies an additional power of $(\xx-\yy)$, so that the difference, of order $\epsilon^{3/2}$, can be neglected. (This observation also lies behind our remark in Sec.~\ref{sec:propagator} that $V$ can be evaluated anywhere along the line between $\xx$ and $\yy$. In Eq.~(\ref{eq:Gfull}) a similar freedom exists for $V$'s argument.)

\section{Discussion}

The beauty and efficacy of the path integral come at a price: ordinary calculus can no longer be taken for granted. As for the Wiener integral used for Brownian motion, the ``paths'' one sums over are rough. They are nowhere differentiable and if one allows time differences on a trajectory to go to zero, the corresponding spatial differences go to zero far more slowly, to wit, $\Delta x \sim \sqrt{\Delta t}$. For this reason, when evaluating the action on broken line paths when a magnetic field is present, the term $\Delta \xx \cdot \bm A$ must be handled carefully. Feynman\cite{feynman} introduced the midpoint rule, namely the use of $(\xx-\yy)\cdot \bm A[(\xx+\yy)/2]$.

In this article we have shown how this prescription arises from a derivation of the path integral that begins from the Hamiltonian and uses a slightly more accurate form of the Trotter product formula. Similar arguments can be made for path integrals on curved spaces,\cite{sensitive} where the need for accuracy is even greater.\cite{greater}

We emphasize that this feature of the path integral is not some minor annoyance, an inessential complication. It is a central feature of quantum mechanics, evident for example in the fact that \textit{velocity} cannot be defined for quantum systems (at least not as the limit of $\Delta x/\Delta t$). Moreover, it plays a central role in applications of the Wiener integral, for example, in the derivation of the Black-Scholes formula.\cite{hull}

\section*{Acknowledgments} This work was supported by the United States National Science Foundation Grant PHY~00~99471 and by the Czech grant ME587.


\begin{thebibliography}{12}
\bibitem[{exp()}]{exponentisbetter}
\bibinfo{note}{If we let $P(\lambda)=\exp(\lambda A)$, $Q(\lambda)=\exp(\lambda
  B)$, $R(\lambda)=\exp(\lambda(A+B))$, the error term can be written $W\equiv
  Q(-\lambda)P(-\lambda)R(\lambda)$. If $\lambda$ is real and $A$ and $B$
  anti-Hermitian (and mathematical pathologies do not intervene) all factors
  are unitary, and therefore so is $W$. One can write the error term (in
  Eq.~(\ref{eq:trotter})) as $PQ(W-1)$. When $W$ is expanded one gets a term
  $\lambda^2 [B,A]/2$, which can be tricky because of the possible
  unboundedness of $A$ and $B$. Note that before the expansion one had a
  product of terms, all of whose norms can be bounded. Nevertheless, in the
  actual proofs one does work with the operators $A$ and $B$. See
  Refs.\cite{nelson} or\cite{lsspibook}.}
\bibitem[{ano()}]{anotherproof}
\bibinfo{note}{If you prefer not to work out long products,
  Eq.~(\ref{eq:betterproduct}) can be derived from previous results.
  Eq.~(\ref{eq:trotterzero}) can also be written with the O($\lambda^2$)
  commutator on the left, but with opposite sign. Now use
  Eq.~(\ref{eq:trotterzero}) as written, but with $\lambda/2$, followed by the
  same thing with the commutator on the left (and with opposite sign), also for
  $\lambda/2$. The commutators now cancel and one obtains, after expanding
  O($\lambda^3$) terms, Eq.~(\ref{eq:betterproduct}), with the roles of $A$ and
  $B$ reversed.}
\bibitem[{gre()}]{greater}
\bibinfo{note}{For the vector potential term, $\bm A$ multiplies $\Delta \xx$,
  which is O($\sqrt{\epsilon}$). On curved spaces, the metric function $g$
  multiplies $(\Delta \xx)^2/\Delta t$, which is O(1), hence yet more stringent
  demands arise for the accurate evaluation of $g$.}
\end{thebibliography}

\end{document}